\documentclass[journal]{IEEEtran}

\usepackage{amsmath, amssymb, amsfonts, bm}

\usepackage[version=4]{mhchem}
\usepackage{chemfig}

\usepackage{algorithm}
\usepackage{algorithmic}

\usepackage{graphicx}
\usepackage{subfigure}
\usepackage{svg}
\usepackage{tikz}
\usetikzlibrary{arrows.meta, decorations.markings, circuits}
\usepackage[american]{circuitikz}

\usepackage{booktabs}
\usepackage{makecell}

\usepackage{textcomp}
\usepackage{soul}       
\usepackage{verbatim}
\usepackage{xcolor}
\usepackage{bbm}

\usepackage{cite}
\usepackage[hidelinks]{hyperref}
\usepackage[capitalise,noabbrev]{cleveref}
\Crefformat{figure}{#2Fig.~#1#3}
\Crefformat{equation}{#2Eq.~(#1)#3}
\crefname{equation}{Eq.}{Eqs.}
\crefname{figure}{Fig.}{Figs.}
\usepackage[
  range-phrase=--,
  per-mode=symbol-or-fraction,
  range-units=single,
  list-units=single,
  detect-all,
  list-final-separator={, and },
]{siunitx}

\usepackage{acro}
\usepackage{mfirstuc}
\DeclareAcronym{MC}{short=MC, long=Molecular Communication}
\DeclareAcronym{MCvD}{short=MCvD, long=Molecular Communication via Diffusion}
\DeclareAcronym{IM}{short=IM, long=information molecule, short-plural=s ,long-plural=s}
\DeclareAcronym{TX}{short=TX, long=transmitter, short-plural=s ,long-plural=s}
\DeclareAcronym{RX}{short=RX, long=receiver, short-plural=s ,long-plural=s}
\DeclareAcronym{SP}{short=SP, long=short pulse}
\DeclareAcronym{LP}{short=LP, long=long pulse}

\DeclareAcronym{ISI}{short=ISI, long=inter-symbol interference}
\DeclareAcronym{BER}{short=BER, long=bit error ratio}
\DeclareAcronym{SER}{short=SER, long=symbol error ratio}
\DeclareAcronym{LTI}{short=LTI, long=linear time-invariant}
\DeclareAcronym{OOK}{short=OOK, long=on-off keying}
\DeclareAcronym{iid}{short=i.i.d., long=independent and identically distributed}

\usepackage{standalone}

\usepackage{titlesec}
\titlespacing\section{0pt}{6pt plus 2pt minus 2pt}{4pt plus 2pt minus 2pt}
\titlespacing\subsection{0pt}{4pt plus 2pt minus 2pt}{3pt plus 2pt minus 2pt}

\ifCLASSINFOpdf
\else
\fi

\begin{document}

\bstctlcite{IEEEexample:BSTcontrol}

\title{Modulation, ISI, and Detection for Langmuir Adsorption-Based Microfluidic Molecular Communication}

\author{Ruifeng~Zheng,~\IEEEmembership{Graduate Student Member,~IEEE}, Pengjie Zhou, Pit Hofmann,~\IEEEmembership{Graduate Student Member,~IEEE}, Martín Schottlender, Fatima Rani,~\IEEEmembership{Member,~IEEE}, Juan A. Cabrera, and Frank H.\,P. Fitzek,~\IEEEmembership{Fellow,~IEEE}
\vspace{-0.3cm}
\thanks{A preliminary version of the receiver model appears in~\cite{zheng2025molecular}.}
\thanks{R.~Zheng, P.~Zhou, P.~Hofmann, M.~Schottlender, F.~Rani, J.\,A.~Cabrera, and F.\,H.\,P.~Fitzek are with the Deutsche Telekom Chair of Communication Networks, Dresden University of Technology, Germany; P.~Hofmann, J.~Cabrera, and F.~Fitzek are also with the Centre for Tactile Internet with Human-in-the-Loop (CeTI), Dresden, Germany, email: \{ruifeng.zheng, pengjie.zhou, pit.hofmann, martin.schottlender, fatima.rani, juan.cabrera, frank.fitzek\}@tu-dresden.de. \\
This work was funded by the German Research Foundation (DFG, Deutsche Forschungsgemeinschaft) as part of Germany’s Excellence Strategy – EXC 2050/2 – Project ID 390696704 – Cluster of Excellence “Centre for Tactile Internet with Human-in-the-Loop” (CeTI) of Dresden University of Technology.
The authors also acknowledge the financial support by the Federal Ministry of Research, Technology and Space (BMFTR) of Germany in the program “Souverän. Digital. Vernetzt.” Joint project 6G-life, project identification number 16KIS2413K, and the program “Verbundprojekt: Disruptive Kommunikationsparadigmen für technologische Souveränität, Resilienz und Shared Prosperity - Translation in Industrie und Aufbau innovativer Technologiedemonstratoren - CommUnity,” project identification number 16KISS012K.
Furthermore, this work was partly supported by the projects IoBNT (grant number 16KIS1994) and MoMiKoSy (Software Campus, grant number 16S23070), funded by the Federal Ministry of Research, Technology and Space (BMFTR).
}}

\maketitle

\begin{abstract}
This paper studies microfluidic molecular communication receivers with finite-capacity Langmuir adsorption driven by an effective surface concentration. In the reaction-limited regime, we derive a closed-form single-pulse response kernel and a symbol-rate recursion for on-off keying that explicitly exposes channel memory and inter-symbol interference. We further develop short-pulse and long-pulse approximations, revealing an interference asymmetry in the long-pulse regime due to saturation. To account for stochasticity, we adopt a finite-receptor binomial counting model, employ pulse-end sampling, and propose a low-complexity midpoint-threshold detector that reduces to a fixed threshold when interference is negligible. Numerical results corroborate the proposed characterization and quantify detection performance versus pulse and symbol durations.
\end{abstract}

\begin{IEEEkeywords}
Molecular communication, Langmuir adsorption, microfluidics, biosensor receiver, binomial counting noise, threshold detection.
\end{IEEEkeywords}

\IEEEpeerreviewmaketitle

\section{Introduction}
\label{sec:introduction}

\IEEEPARstart{M}{icrofluidic} molecular communication (MC) with surface-based biosensors is a promising paradigm for enabling reliable information transfer in lab-on-a-chip platforms and bio-nanotechnological systems, where \acp{IM} are transported by laminar flow and detected via reversible binding to receptors on a sensing surface~\cite{hassibi2009real,squires2008making,xu2017real}. Such receivers are attractive for applications including in situ biochemical sensing, closed-loop synthetic biology, and microfluidic diagnostics, as they can directly interface chemical signals with electronic readout circuits (e.g., field-effect transistor biosensors)~\cite{kuscu2018modeling,kuscu2021fabrication,scherer2025closed}. A central challenge, however, lies in accurately characterizing the \emph{receiver dynamics} under realistic biochemical constraints. In particular, finite receptor capacity and reversible binding/unbinding kinetics introduce nonlinear saturation and a persistent temporal memory~\cite{xu2017real,squires2008making,zheng2025molecular}, which shape \ac{ISI} and fundamentally impact detection performance~\cite{deng2015modeling,kuscu2018modeling,zheng2025anis,zheng2025molecular}.

Many MC studies model the receiver as an instantaneous detector once IMs reach the sensing surface~\cite{yilmaz2014three,deng2015modeling,huang2019spatial,zheng2020noise}. In contrast, practical surface-based biosensors often operate in a \emph{reaction-limited} regime, where binding/unbinding kinetics dominate the sensing timescale and introduce latency and temporal memory~\cite{squires2008making}. Existing receiver models span passive concentration sensing and active boundary interactions (e.g., absorption or reversible adsorption)~\cite{farsad2016comprehensive,guo2016molecular,yilmaz2014three,deng2015modeling}. For microfluidic biosensing, a convection--diffusion--reaction framework with finite reactive surfaces was developed in~\cite{kuscu2018modeling}, and graphene field-effect transistor (GFET)-based DNA biosensor receivers were experimentally demonstrated in~\cite{kuscu2021fabrication}. Despite these advances, a communication-oriented \emph{symbol-level} characterization that reveals how finite-receptor Langmuir dynamics induce \ac{ISI} under \ac{OOK} and enables low-complexity detection and error-performance evaluation under counting noise remains needed.

Motivated by this need, we consider \ac{OOK} signaling in which mass transport is abstracted into a time-varying \emph{effective surface concentration} driving Langmuir adsorption at the receiver surface. The finite-receptor reaction dynamics govern how quickly the receiver output rises during molecule injection and decays after injection stops, yielding distinct short-pulse (SP) and long-pulse (LP) behaviors. A preliminary receiver modeling study appears in~\cite{zheng2025molecular}. Here, we substantially extend it to an end-to-end symbol-rate communication framework by incorporating modulation, ISI analysis, and detection. This work is also complementary to our Markov-based framework for DNA-based MC~\cite{zheng2025DNA-Based}, focusing on adsorption-based biosensor receivers and their implications for symbol-level performance.

The main contributions of this paper are summarized as follows: (1) We model microfluidic transport via an effective surface concentration driving finite-capacity Langmuir adsorption and derive a symbol-rate recursion that exposes channel memory and \ac{ISI}; (2) we derive a closed-form single-pulse response kernel and develop short-/long-pulse (SP/LP) approximations for characterizing the symbol-rate response and \ac{ISI}; and (3) we incorporate finite-receptor binomial counting noise, adopt pulse-end sampling, and propose a low-complexity midpoint-threshold detector for performance evaluation.

\section{System Model}
\label{sec:system_model}

\begin{figure}[t]
    \centering
    \includegraphics[width=1.0\linewidth]{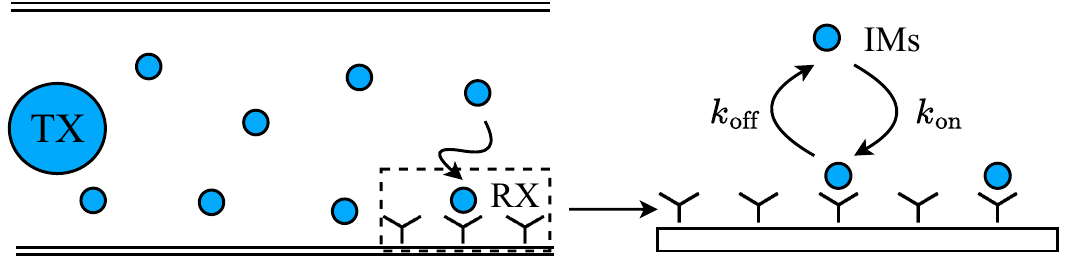}
    \vspace{-0.3cm}
    \caption{Schematic illustration of reversible binding of IMs at the receiver surface.}
    \label{fig:schematic}
    \vspace{-0.3cm}
\end{figure}

In this section, we describe the considered microfluidic MC system with a surface-based biosensor receiver, including the transport-driven effective surface concentration model, the \ac{OOK} modulation scheme, and the resulting Langmuir adsorption dynamics. We then derive the single-pulse response kernel and use it to characterize the deterministic response to an \ac{OOK} symbol sequence, highlighting the origin of \ac{ISI} and providing SP and LP approximations that will be used later for performance analysis.

\subsection{System Overview}
We consider a microfluidic conduit that releases \acp{IM} from the transmitter to a surface-based biosensor receiver with $N_p$ identical, non-interacting binding sites, see~\cref{fig:schematic}. Under stable laminar flow and diffusion-assisted mixing, the transport effects near the sensing surface are captured by a time-varying \emph{effective surface concentration} $c(t)$, consistent with~\cite{squires2008making,hassibi2009real}. In this paper, we treat $c(t)$ as the effective channel input; the mapping from the injected waveform to $c(t)$ can be obtained from a transport model and is established in our prior work~\cite{zheng2025molecular}.

Under the \emph{reaction-limited} assumption, the mean number of occupied sites $N_b(t)\in[0,N_p]$ evolves according to the Langmuir kinetics
\begin{equation}
\frac{dN_b(t)}{dt}
= k_{\rm on}c(t)\bigl(N_p-N_b(t)\bigr) - k_{\rm off}N_b(t),
\label{eq:langmuir_ode}
\end{equation}
where $k_{\rm on}$ and $k_{\rm off}$ are the binding and unbinding rate constants, respectively.

\subsection{Transmitter Modulation}
\label{subsec:modulation}

We consider \ac{OOK} modulation with a symbol interval $T_b$ and pulse duration $T\le T_b$. The symbol $1$ is represented by a rectangular concentration pulse of amplitude $c_0$ and duration $T$, while the symbol $0$ corresponds to no pulse. For a symbol sequence $\{a_i\}_{i=0}^{K-1}$, $a_i\in\{0,1\}$, let $t_i\triangleq iT_b$ denote the start time of the $i$-th symbol. The resulting effective surface concentration is
\begin{equation}
x(t)\triangleq c(t)
= c_0\sum_{i=0}^{K-1} a_i\Big[u(t-t_i)-u\bigl(t-t_i-T\bigr)\Big],
\label{eq:input_sequence}
\end{equation}
where $u(t)$ is the unit-step function. For a single bit-$1$ pulse starting at $t_0=0$, \cref{eq:input_sequence} reduces to $x(t)=c_0[u(t)-u(t-T)]$.

\subsection{Single-Pulse Response Kernel}
\label{subsec:kernel}

\begin{figure}[t]
    \centering
    \subfigure[$T=0.5$ min, $\tau_{\rm off}=10$ min\label{fig:h_a}]{%
        \includegraphics[width=0.49\linewidth]{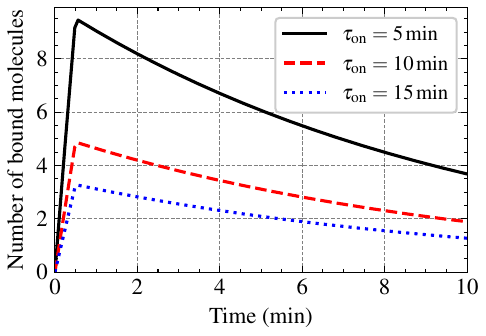}}\hfill
    \subfigure[$T=0.5$ min, $\tau_{\rm on}=10$ min\label{fig:h_b}]{%
        \includegraphics[width=0.49\linewidth]{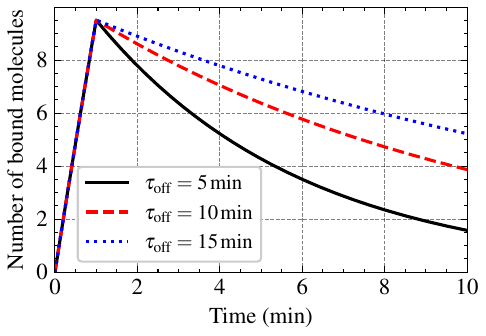}}\\[-1mm]
    \subfigure[$T=100$ min, $\tau_{\rm off}=10$ min\label{fig:h_c}]{%
        \includegraphics[width=0.49\linewidth]{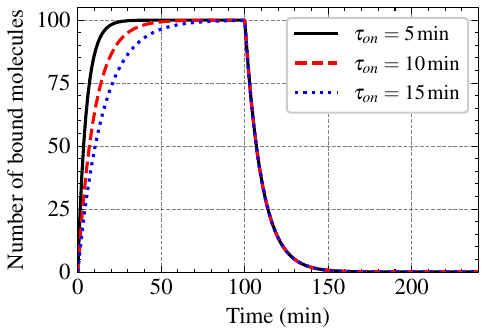}}\hfill
    \subfigure[$T=100$ min, $\tau_{\rm on}=10$ min\label{fig:h_d}]{%
        \includegraphics[width=0.49\linewidth]{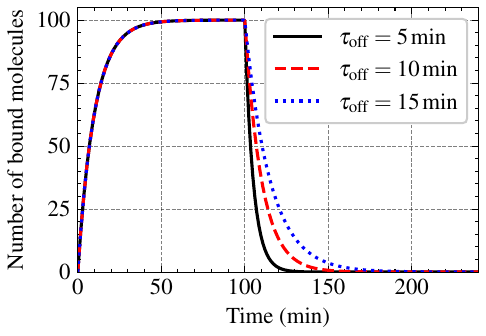}}
    \caption{Single-pulse response kernel $h_T(t)$ in the SP and LP regimes, illustrating the impact of $\tau_{\rm on}$ on the rise phase and $\tau_{\rm off}$ on the post-pulse decay (temporal memory).}
    \label{fig:hT}
    \vspace{-0.3cm}
\end{figure}

We define the \emph{single-pulse response kernel} $h_T(t)$ as the zero-state response (i.e., $N_b(0)=0$) to a single rectangular pulse in~\cref{eq:input_sequence}. Solving~\cref{eq:langmuir_ode} yields
\begin{equation}
h_T(t)=
\begin{cases}
0, & t\le 0,\\[2pt]
N_b^{\infty}\!\left[1-\exp\!\left(-t/\tau_{\mathrm{on}}\right)\right], & 0<t\le T,\\[4pt]
N_b^{\mathrm{end}}\exp\!\left[-(t-T)/\tau_{\mathrm{off}}\right], & t>T,
\end{cases}
\label{eq:hT}
\end{equation}
where the steady bound count and the characteristic binding/unbinding time constants are
\begin{equation}
N_b^{\infty} = N_p \frac{c_0 / K_\mathrm{D}}{1 + c_0 / K_\mathrm{D}}, \quad
\tau_{\mathrm{on}} = \frac{1}{k_{\mathrm{on}} c_0 + k_{\mathrm{off}}}, \quad
\tau_{\mathrm{off}} = \frac{1}{k_{\mathrm{off}}},
\label{eq:NbInf_taus}
\end{equation}
with dissociation constant $K_\mathrm{D}=k_{\mathrm{off}}/k_{\mathrm{on}}$. Moreover, the bound count at the end of the concentration pulse is
\begin{equation}
N_b^{\mathrm{end}}\triangleq h_T(T)
= N_b^{\infty}\!\left[1-\exp\!\left(-T/\tau_{\mathrm{on}}\right)\right],
\label{eq:nb_end}
\end{equation}
which is the pulse-end (and peak) value of $h_T(t)$. The time constants $\tau_{\mathrm{on}}$ and $\tau_{\mathrm{off}}$ characterize the rise and decay dynamics of the receiver response, respectively; in particular, $\tau_{\mathrm{off}}$ governs the post-pulse decay and thus the effective temporal memory. This temporal memory implies that residual bound molecules remain after the pulse ends and interfere with subsequent symbols, thereby giving rise to \ac{ISI}.

In the SP regime, we approximate the exponential rise by its first-order expansion. Specifically, for $T\le 0.2\,\tau_{\mathrm{on}}$, the kernel in~\cref{eq:hT} can be approximated as
\begin{equation}
h_{\mathrm{SP}}(t)\approx
\begin{cases}
0, & t\le 0,\\[2pt]
N_b^{\infty}\dfrac{t}{\tau_{\mathrm{on}}}, & 0<t\le T,\\[6pt]
N_b^{\infty}\dfrac{T}{\tau_{\mathrm{on}}}\exp\!\left[-(t-T)/\tau_{\mathrm{off}}\right], & t>T.
\end{cases}
\label{eq:hT_SP}
\end{equation}

In the LP regime, the response approaches the steady-state within the pulse. In particular, for $T\ge 5\,\tau_{\mathrm{on}}$ we have $\exp(-T/\tau_{\mathrm{on}})\le e^{-5}$ and thus $N_b^{\mathrm{end}}\approx N_b^{\infty}$; the initial rise transient of duration is negligible relative to the pulse duration. Therefore, we adopt the LP approximation
\begin{equation}
h_{\mathrm{LP}}(t)\approx
\begin{cases}
0, & t\le 0,\\[2pt]
N_b^{\infty}, & 0<t\le T,\\[4pt]
N_b^{\infty}\exp\!\bigl[-(t-T)/\tau_{\mathrm{off}}\bigr], & t>T.
\end{cases}
\label{eq:hT_LP}
\end{equation}
\Cref{fig:hT} illustrates $h_T(t)$ in the SP and LP regimes. In \cref{fig:h_a,fig:h_c}, varying $\tau_{\rm on}$ mainly changes the rise behavior within the pulse, whereas in \cref{fig:h_b,fig:h_d}, varying $\tau_{\rm off}$ directly controls the post-pulse decay tail. A larger $\tau_{\rm off}$ yields a slower decay and a longer \emph{temporal memory}, meaning that the response to a symbol persists into subsequent symbol intervals and becomes the primary physical source of \ac{ISI} in this work.

\subsection{Deterministic Response of OOK Sequence}
\label{subsec:det_response}

We characterize the deterministic (mean) binding response $y(t)\triangleq N_b(t)$ to an \ac{OOK} symbol sequence driven by the effective surface concentration $c(t)$ in~\cref{eq:input_sequence}. Let $t_i\triangleq iT_b$ for $i=0,1,\ldots,K-1$, define the symbol-start state $Y_i\triangleq y(t_i)$, and denote the pulse-end sampling time by
\begin{equation}
t_{s,i}\triangleq t_i+T.
\label{eq:tsi}
\end{equation}
Substituting~\cref{eq:input_sequence} into~\cref{eq:langmuir_ode}, the exact response within the $i$-th symbol interval $t\in[t_i,t_{i+1})$ (where $t_{i+1}=t_i+T_b$) is given as follows. 

For $a_i=1$,
\begin{equation}
y(t)=
\begin{cases}
\begin{array}{@{}l@{\;\;}l@{}}
N_b^{\infty}
+\bigl(Y_i-N_b^{\infty}\bigr)\exp\!\left(-\dfrac{t-t_i}{\tau_{\rm on}}\right),
& t\in[t_i,t_{s,i}), \\[6pt]
y(t_{s,i})\exp\!\left(-\dfrac{t-t_{s,i}}{\tau_{\rm off}}\right),
& t\in[t_{s,i},t_{i+1}),
\end{array}
\end{cases}
\label{eq:piecewise_ai1}
\end{equation}
where the pulse-end response is
\begin{equation}
y(t_{s,i})
= N_b^{\infty}+\bigl(Y_i-N_b^{\infty}\bigr)\exp\!\left(-T/\tau_{\rm on}\right).
\label{eq:ytsi}
\end{equation}

For $a_i=0$,
\begin{equation}
y(t)=Y_i\exp\!\left(-(t-t_i)/\tau_{\rm off}\right),
\qquad t\in[t_i,t_{i+1}).
\label{eq:piecewise_ai0}
\end{equation}

Evaluating at $t=t_{i+1}$ yields the closed-form recursion
\begin{equation}
Y_{i+1}=
\begin{cases}
y(t_{s,i})\exp\!\left(-(T_b-T)/\tau_{\rm off}\right), & a_i=1,\\[2pt]
Y_i\exp\!\left(-T_b/\tau_{\rm off}\right), & a_i=0,
\end{cases}
\label{eq:Y_recursion}
\end{equation}
which explicitly reveals how past symbols determine the evolution of the symbol-start state sequence $\{Y_i\}$.

\subsection{Kernel-Based SP/LP Approximations}
\label{subsec:kernel_approx}

In the SP regime, we employ the SP kernel approximation in~\cref{eq:hT_SP}. When the occupancy remains low (so that the Langmuir dynamics can be well approximated by a linearization around the operating point), the overall response can be approximated by a superposition of shifted single-pulse kernels, i.e.,
\begin{equation}
y_{\rm SP}(t)\approx \sum_{i=0}^{K-1} a_i\,h_{\rm SP}(t-iT_b),
\label{eq:SP_approx}
\end{equation}
where $h_{\rm SP}(t)$ is given in~\cref{eq:hT_SP}.

In the LP regime, the binding response approaches steady state within the pulse; 
for $T\ge 5\,\tau_{\rm on}$ we have $N_b^{\mathrm{end}}\approx N_b^{\infty}$ and thus the LP kernel approximation in~\cref{eq:hT_LP} applies. Since saturation makes the system generally not \ac{LTI}, the following expression should be understood as a pulse-wise benchmark approximation. When \ac{ISI} is negligible (e.g., sufficiently large $T_b$), we use
\begin{equation}
y_{\rm LP}(t)\approx \sum_{i=0}^{K-1} a_i\,h_{\rm LP}(t-iT_b),
\label{eq:LP_approx}
\end{equation}
where $h_{\rm LP}(t)$ is given in~\cref{eq:hT_LP}.

\begin{figure}[t]
    \centering
    \subfigure[SP input\label{fig:example_a}]{%
        \includegraphics[width=0.49\linewidth]{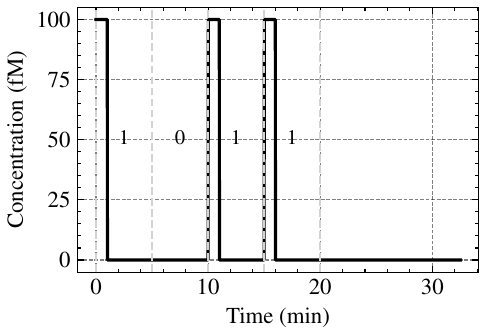}}\hfill
    \subfigure[LP input\label{fig:example_b}]{%
        \includegraphics[width=0.49\linewidth]{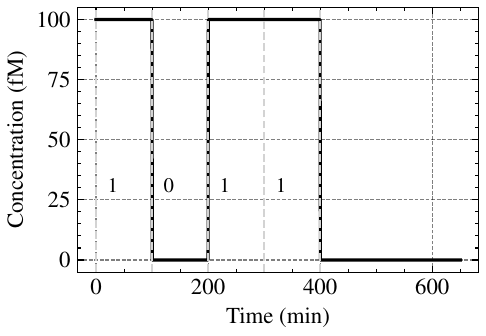}}\\[-1mm]
    \subfigure[SP response\label{fig:example_c}]{%
        \includegraphics[width=0.49\linewidth]{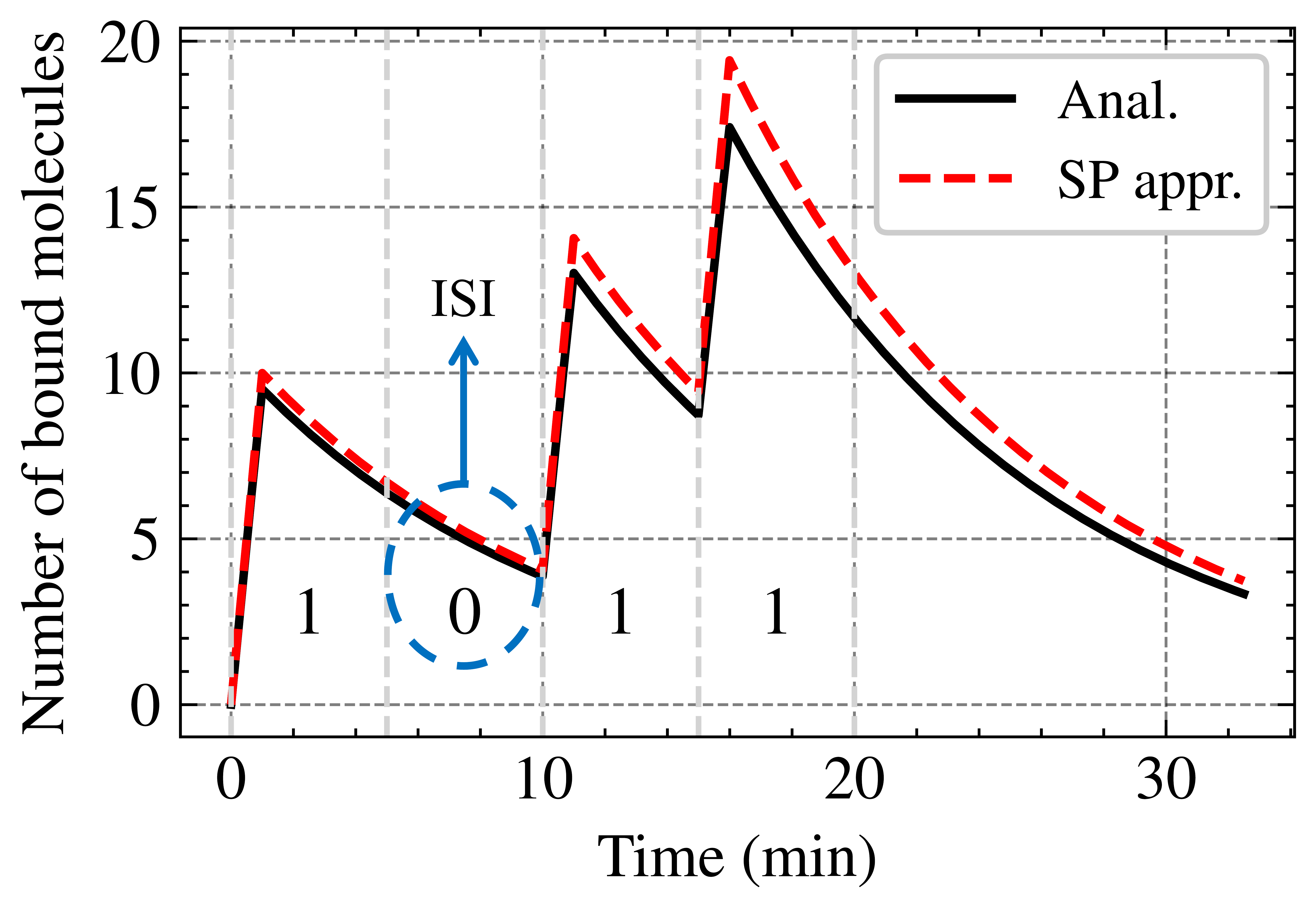}}\hfill
    \subfigure[LP response\label{fig:example_d}]{%
        \includegraphics[width=0.49\linewidth]{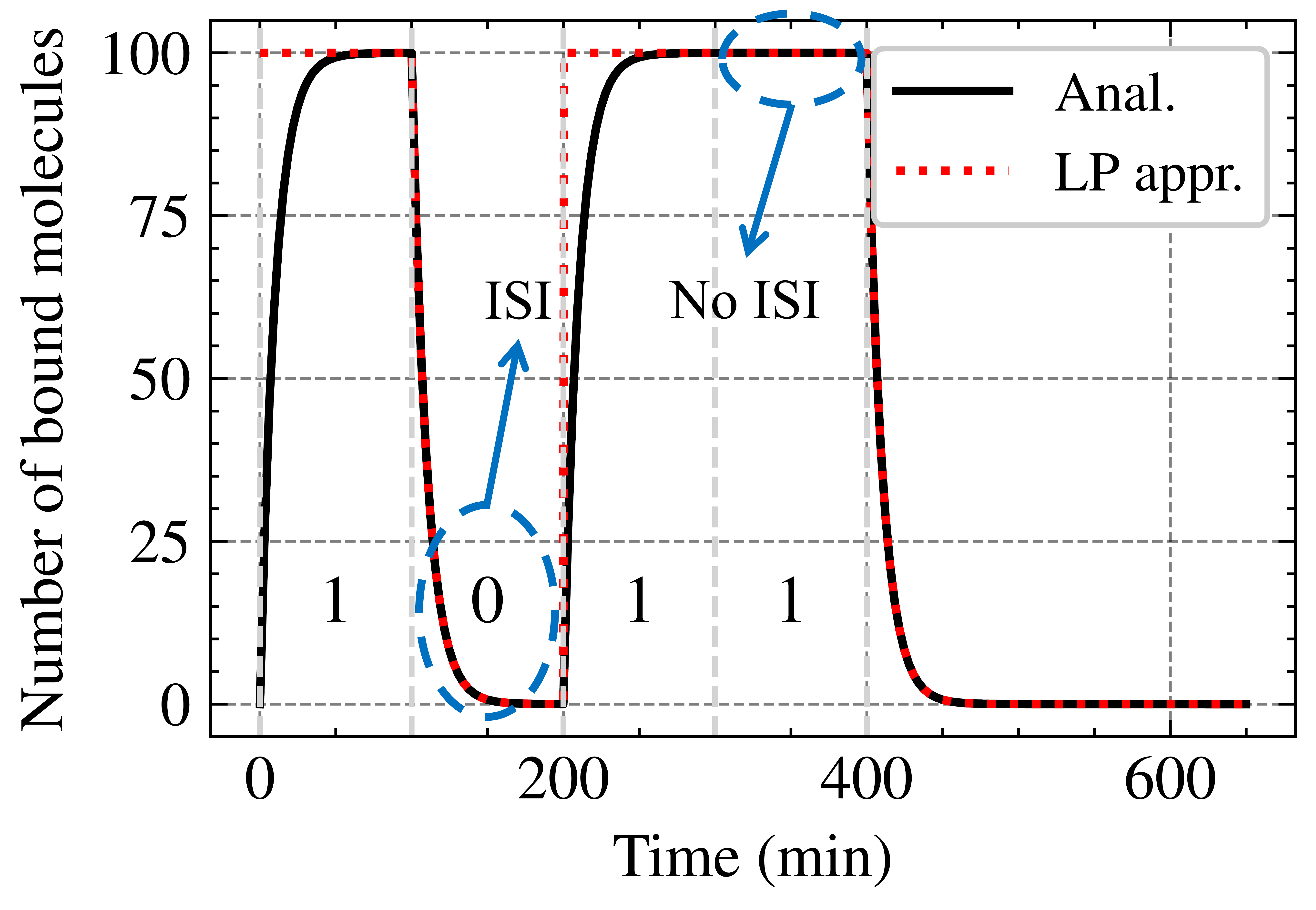}}
    \caption{Illustration of OOK modulation and deterministic receiver response for the symbol sequence $[1,0,1,1]$ in the SP ($T=0.1\,\tau_{\rm on}$, $T_b=0.5\,\tau_{\rm on}$) and LP ($T=10\,\tau_{\rm on}$, $T_b=10\,\tau_{\rm on}$) regimes. (a),(c) correspond to SP signaling and compare the analytical response with the SP approximation; (b),(d) correspond to LP signaling and compare the analytical response with the LP approximation. Parameters: $\tau_{\rm on}=10$~min and $\tau_{\rm off}=10$~min.}
    \label{fig:example}
    \vspace{-0.3cm}
\end{figure}

\subsection{ISI Mechanism and Asymmetry}
\label{subsec:isi_influence}

\Cref{fig:example} compares SP and LP signaling for the same symbol sequence. In the SP case (\cref{fig:example_c}), the receiver operates in a low-occupancy regime, where the adsorption dynamics are approximately linear, and the response resembles that of classical passive-receiver models: \ac{ISI} is primarily induced by the exponential dissociation (unbinding) tail of previously bound molecules, governed by $\tau_{\rm off}$. In the LP case (\cref{fig:example_d}), receptor saturation introduces a pronounced asymmetry. For $1\!\to\!0$ transitions, residual bound molecules after pulse end yield appreciable \ac{ISI}. In contrast, for $1\!\to\!1$ transitions, the bound count is already close to saturation (limited by the finite binding-site capacity $N_p$), so the additional \ac{ISI} contribution is small and the resulting \ac{ISI} is substantially reduced. This asymmetry will be reflected in the detection and error-performance results in later sections.


\section{Noisy Observation Model and Detection}
\label{sec:noise_detection}
\begin{table}[tbp] 
\centering
\caption{Simulation Parameters.}
\label{tab:parameters}
\renewcommand{\arraystretch}{1.2}
\begin{tabular}{@{}lccc@{}}
\toprule
\textbf{Parameter} & \textbf{Symbol} & \textbf{Value} & \textbf{Unit} \\ \midrule
Number of transmitted symbols & $K$ & $10^5$ & -- \\
Number of binding sites & $N_p$ & $10^6$ & -- \\
Binding rate constant & $k_\text{on}$ & $10^8$ & $\mathrm{M^{-1}min^{-1}}$ \\
Unbinding rate constant & $k_\text{off}$ & 0.1 & $\mathrm{min^{-1}}$ \\
Pulse concentration & $c_0$ & 100 & $\mathrm{fM}$ \\
\bottomrule
\end{tabular}
\vspace{-0.3cm}
\end{table}

In this section, we incorporate receiver noise and develop a low-complexity detection scheme. Specifically, we model the observed bound count using a finite-receptor binomial counting model, define a pulse-end sampling statistic, and apply a simple midpoint-threshold detector. For the special case of negligible ISI, the detector further reduces to a fixed threshold.

\subsection{Noisy Observation Model}
\label{subsec:noise_model}

The receiver output $y(t)\triangleq N_b(t)$ in~\cref{eq:piecewise_ai1,eq:piecewise_ai0} represents the deterministic (mean) number of occupied binding sites. Since binding and unbinding are stochastic events over a finite population of $N_p$ sites, we model the measured bound count as a binomial random variable. Let $\tilde N_b(t)$ denote the observed number of occupied sites at time $t$. Conditioned on the mean occupancy probability $p(t)\triangleq y(t)/N_p$, we assume
\begin{equation}
\tilde N_b(t)\,\big|\,y(t)\ \sim\ \mathcal{B}\!\left(N_p,\,p(t)\right),
\label{eq:binom_model}
\end{equation}
where $\mathcal{B}(n,p)$ denotes a binomial distribution with $n$ independent trials and success probability $p$.

For analytical tractability, we also employ a Poisson approximation. When $N_p$ is large and $p(t)$ is small such that $N_p p(t)=y(t)$ remains finite, the conditional distribution in~\cref{eq:binom_model} can be approximated as
\begin{equation}
\tilde N_b(t)\,\big|\,y(t)\ \approx\ \mathcal{P}\!\left(y(t)\right),
\label{eq:poisson_model}
\end{equation}
where $\mathcal{P}(\lambda)$ denotes a Poisson distribution with mean $\lambda$.

\subsection{Sampling and Decision Statistic}
\label{subsec:sampling_stat}

We sample the receiver output at the pulse-end time $t_{s,i}$ defined in~\cref{eq:tsi}, and define the decision statistic
\begin{equation}
z_i \triangleq \tilde N_b(t_{s,i}).
\label{eq:zi}
\end{equation}
Let $\mu_{m,i}\triangleq \mathbb{E}\{z_i\mid \mathcal{H}_m\}$ denote the conditional mean under hypothesis $\mathcal{H}_m$ ($m\in\{0,1\}$, corresponding to $a_i=m$). From~\cref{eq:piecewise_ai1,eq:piecewise_ai0}, we obtain
\begin{align}
\mu_{1,i}
&= y(t_{s,i})
= N_b^{\infty}+\bigl(Y_i-N_b^{\infty}\bigr)\exp\!\left(-T/\tau_{\rm on}\right),
\label{eq:mu1}\\
\mu_{0,i}
&= y(t_{s,i})
= Y_i\exp\!\left(-T/\tau_{\rm off}\right),
\label{eq:mu0}
\end{align}
where $Y_i\triangleq y(t_i)$ is the symbol-start state. Accordingly, under the binomial counting model in~\cref{eq:binom_model},
\begin{equation}
z_i\mid \mathcal{H}_m \sim \mathcal{B}\!\left(N_p,\,\mu_{m,i}/N_p\right),\quad m\in\{0,1\}.
\label{eq:zi_cond}
\end{equation}

\subsection{Midpoint-Threshold Detector}
\label{subsec:detector}

To keep the detection rule simple, we adopt an integer-valued midpoint threshold
\begin{equation}
\eta_i \triangleq \left\lfloor \frac{\mu_{0,i}+\mu_{1,i}}{2}\right\rceil,
\label{eq:eta_mid}
\end{equation}
and decide
\begin{equation}
\hat a_i=
\begin{cases}
1, & z_i>\eta_i,\\
0, & z_i\le \eta_i,
\end{cases}
\label{eq:decision}
\end{equation}
where $\lfloor\cdot\rceil$ denotes rounding to the nearest integer.

Since $\eta_i$ depends on the symbol-start state $Y_i$, we employ a decision-feedback (DF) state tracker~\cite{belfiore1979decision}. Specifically, the receiver maintains a symbol-rate state estimate $\hat Y_i$, initializes $\hat Y_0=0$, and computes $\eta_i$ by substituting $\hat Y_i$ into~\cref{eq:mu0,eq:mu1}. After deciding $\hat a_i$ using~\cref{eq:decision}, the state estimate is updated via the same recursion as~\cref{eq:Y_recursion} with $a_i$ replaced by $\hat a_i$, i.e.,
\begin{equation}
\hat Y_{i+1}=
\begin{cases}
\mu_{1}(\hat Y_i)\exp\!\left[-(T_b-T)/\tau_{\mathrm{off}}\right], & \hat a_i=1,\\
\hat Y_i\exp\!\left(-T_b/\tau_{\mathrm{off}}\right), & \hat a_i=0,
\end{cases}
\label{eq:df_recursion}
\end{equation}
where $\mu_{1}(\hat Y_i)\triangleq N_b^{\infty}+(\hat Y_i-N_b^{\infty})\exp(-T/\tau_{\mathrm{on}})$.

When \ac{ISI} is negligible (e.g., $T_b$ is large enough that the residual occupancy decays to $Y_i \approx 0$ at the beginning of symbol~$i$), the conditional means satisfy $\mu_{0,i}\approx 0$ and $\mu_{1,i}\approx N_b^{\mathrm{end}}$; hence, a fixed midpoint threshold $\eta \approx \lfloor N_b^{\mathrm{end}}/2\rceil$ is sufficient.

\section{Numerical Results}
\label{sec:numerical_results}

In this section, we present numerical evaluations to quantify the bit error ratio (BER) performance of the proposed pulse-end sampling and DF  midpoint-threshold detector under binomial counting noise. All simulations are conducted in MATLAB, with the parameters summarized in~\cref{tab:parameters}, unless otherwise specified.

\subsection{Simulation Setup}
\label{subsec:sim_setup_ber}
\begin{figure*}[t]
    \centering
    \subfigure[Varying $T_b$.]{
        \label{fig:ber_vs_Tb}
        \centering
        \includegraphics[width=0.45\linewidth]{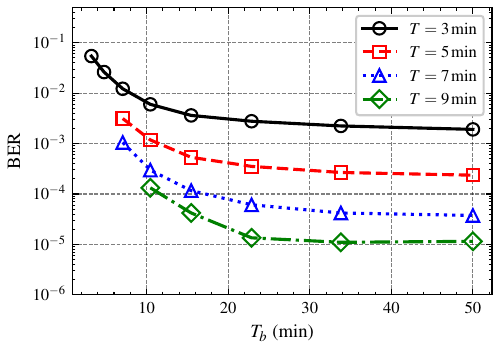}
    }
    \hfill
    \subfigure[Varying $N_p$.]{
        \label{fig:ber_vs_Np}
        \centering
        \includegraphics[width=0.45\linewidth]{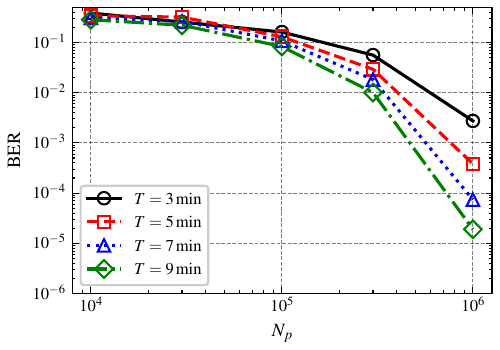}
    }
    \vspace{-0.3cm}
    \caption{BER performance of the DF midpoint-threshold detector. (a) BER versus symbol interval $T_b$ for different pulse durations $T$. (b) BER versus the number of binding sites $N_p$ for different $T$, with $T_b=20~\mathrm{min}$.}
    \label{fig:ber}
    \vspace{-0.3cm}
\end{figure*}
We consider \ac{OOK} modulation with pulse duration $T$ and symbol interval $T_b$. For each configuration, we generate $K$ \ac{iid} equiprobable symbols and compute the mean state evolution via the recursion in~\cref{eq:Y_recursion}. Noisy pulse-end samples $z_i=\tilde N_b(t_{s,i})$ with $t_{s,i}=t_i+T$ are drawn from the binomial counting model in~\cref{eq:binom_model}, and symbols are detected using~\cref{eq:eta_mid} with DF state tracking based on~\cref{eq:Y_recursion}. The BER is estimated as
\begin{equation}
\mathrm{BER}\triangleq \frac{1}{K}\sum_{i=0}^{K-1}\mathbbm{1}\{\hat a_i\neq a_i\},
\label{eq:ber_def}
\end{equation}
where $\mathbbm{1}\{\cdot\}$ denotes the indicator function, i.e., $\mathbbm{1}\{A\}=1$ if $A$ is true and $\mathbbm{1}\{A\}=0$ otherwise.

\subsection{Impact of Symbol Interval}
\label{subsec:ber_vs_tb}

In the considered Langmuir adsorption-based microfluidic MC system, the dominant performance limitation is \ac{ISI} induced by the receptor temporal memory. Increasing the symbol interval $T_b$ allows the bound count to relax further between symbols, thereby mitigating \ac{ISI}. \Cref{fig:ber_vs_Tb} shows the BER as a function of the symbol interval $T_b$ for several pulse durations $T$, with the constraint $T\le T_b$. As $T_b$ increases from $3$~min to $50$~min, all BER curves decrease monotonically. This behavior confirms that $T_b$ is the main design parameter controlling \ac{ISI}, since a larger $T_b$ provides more time for the post-pulse dissociation to reduce the residual occupancy before the next symbol starts.

Moreover, the BER reduction becomes marginal once $T_b$ exceeds approximately $30$~min. This saturation is consistent with the temporal-memory decay governed by $\tau_{\rm off}$ (here $\tau_{\rm off}=10$~min). After the pulse ends, the residual response decays exponentially as $\exp(-\Delta/\tau_{\rm off})$, thus, at $\Delta=\tau_{\rm off}$ it drops to $e^{-1}\!\approx\!37\%$ of its pulse end value, at $\Delta=3\tau_{\rm off}$ to $e^{-3}\!\approx\!5\%$, and at $\Delta=5\tau_{\rm off}$ to $e^{-5}\!\approx\!0.7\%$ (see details in~\cite{zheng2025molecular}). Therefore, when $T_b\gtrsim 3\tau_{\rm off}$, the residual occupancy carried into the next symbol is already small
, and further increasing $T_b$ yields only limited additional \ac{ISI} mitigation, resulting in an approximately flat BER.

For a fixed symbol interval $T_b$, the BER decreases as the pulse duration $T$ increases. A longer pulse yields a larger pulse-end bound count $N_b^{\mathrm{end}}$ (cf.~\cref{eq:nb_end}), thereby increasing the sampled mean under $\mathcal{H}_1$ (i.e., $\mu_{1,i}$ in~\cref{eq:mu1}). Consequently, the mean separation between the two hypotheses at the sampling time increases, i.e., $\mu_{1,i}-\mu_{0,i}$ (cf.~\cref{eq:mu0,eq:mu1}), which improves detectability. Under the binomial counting model, a larger mean count yields a more reliable pulse-end sample, resulting in a lower BER.

\subsection{Impact of the Number of Binding Sites}
\label{subsec:ber_vs_np}

\Cref{fig:ber_vs_Np} shows the BER as a function of the number of binding sites $N_p$ for different pulse durations $T$. The symbol interval is fixed to $T_b=20$~min, which corresponds to $2\tau_{\rm off}$ for the considered parameters; thus, the residual response between adjacent symbols is noticeably reduced, and the performance trend is largely dominated by counting noise. As $N_p$ increases from $10^{4}$ to $10^{6}$, the BER decreases for all $T$. This is expected since the steady state bound count scales linearly with the number of available sites, i.e., $N_b^{\infty}\propto N_p$ in~\cref{eq:NbInf_taus}. With a small $N_p$ (e.g., $10^{4}$--$10^{5}$), the pulse-end mean bound count remains low, hence the binomial counting fluctuations are large relative to the mean, leading to unreliable decisions and an unacceptably high BER. As $N_p$ grows (e.g., $10^{5}$--$10^{6}$), the mean count increases while the relative counting uncertainty decreases, resulting in a higher effective SNR of the pulse-end statistic and a substantially lower BER.

For a fixed $N_p$, a larger $T$ further improves performance. A longer pulse yields a larger pulse-end bound count $N_b^{\mathrm{end}}$ (cf.~\cref{eq:nb_end}), which increases the sampled mean under $\mathcal{H}_1$ and enlarges the mean separation between hypotheses (cf.~\cref{eq:mu0,eq:mu1}). Consequently, the impact of counting noise becomes less pronounced, and the BER decreases.

\section{Conclusion}
\label{sec:conclusion}

This paper studied Langmuir adsorption--based microfluidic molecular communication with finite receptor capacity. We derived a closed-form single-pulse response kernel and a symbol-rate recursion for \ac{OOK}, which together reveal the channel memory and the resulting \ac{ISI}. We further developed SP/LP approximations and showed an LP \ac{ISI} asymmetry induced by receptor saturation. To account for stochasticity, we adopted a binomial counting model and proposed a low-complexity pulse-end-sampling midpoint-threshold detector with decision-feedback (DF) state tracking. Numerical results validated the analysis and highlighted the impacts of $T_b$ and $N_p$ on \ac{ISI} and detection reliability. Future work includes joint transport--reaction modeling and optimized detection in the strong-\ac{ISI} regime.

\bibliographystyle{IEEEtran}
\bibliography{references}

\end{document}